\pdfoutput=1
\documentclass[twocolumn,aps,prl,amsmath,amssymb,showpacs,superscriptaddress]{revtex4}

\usepackage{graphicx}
\usepackage{graphics}
\usepackage[latin1]{inputenc}
\usepackage{latexsym}
\usepackage{amsmath}
\usepackage{amssymb}

\def\endproof{\vrule height6pt width6pt depth0pt}

\begin{document}
\title{Necessary and Sufficient Detection Efficiency for the Mermin Inequalities}
\author{Ad{\'a}n Cabello}
\email{adan@us.es}
\affiliation{Departamento de F{\'\i}sica Aplicada II, Universidad de Sevilla, E-41012 Sevilla, Spain}
\author{David Rodr{\'\i}guez}
%\email{davidroba@hotmail.com}
\affiliation{Departamento de
F{\'\i}sica Aplicada III, Universidad de Sevilla, E-41092 Sevilla,
Spain}
\author{Ignacio Villanueva}
%\email{ignaciov@mat.ucm.es}
\affiliation{Departamento de An{\'a}lisis Matem{\'a}tico, Universidad
Complutense, E-28040 Madrid, Spain}

%%%%%%%%%%%%%%%%%%%%%%%%%%%%%%%%%%%%%%%%%%%%%%%%%%%%%%%%%%%%%%%%%%%

\date{\today}
%First version: 26 November 2007
%This version: 09 September 2008, after PRL proofs.

%%%%%%%%%%%%%%%%%%%%%%%%%%%%%%%%%%%%%%%%%%%%%%%%%%%%%%%%%%%%%%%%%%%

\begin{abstract}
We prove that the threshold detection efficiency for a loophole-free
Bell experiment using an $n$-qubit Greenberger-Horne-Zeilinger state
and the correlations appearing in the $n$-partite Mermin inequality
is $n/(2n-2)$. If the detection efficiency is equal to or lower than
this value, there are local hidden variable models that can simulate
all the quantum predictions. If the detection efficiency is above
this value, there is no local hidden variable model that can
simulate all the quantum predictions.
\end{abstract}

%%%%%%%%%%%%%%%%%%%%%%%%%%%%%%%%%%%%%%%%%%%%%%%%%%%%%%%%%%%%%%%%%%%

\pacs{03.65.Ud,
%Entanglement and quantum nonlocality
%(e.g. EPR paradox, Bell's inequalities, GHZ states, etc.)
03.67.Mn,
%Entanglement production, characterization and manipulation,
42.50.Xa}
%Optical tests of quantum theory

\maketitle

%%%%%%%%%%%%%%%%%%%%%%%%%%%%%%%%%%%%%%%%%%%%%%%%%%%%%%%%%%%%%%%%%%%

%\section{Introduction}

%%%%%%%%%%%%%%%%%%%%%%%%%%%%%%%%%%%%%%%%%%%%%%%%%%%%%%%%%%%%%%%%%%%

Quantum nonlocality is the impossibility of reproducing the quantum
correlations between the results of distant measurements using local
hidden variable (LHV) theories. This impossibility is shown either
by the violation of a Bell inequality \cite{Bell64} or the
impossibility of ascribing predefined results which simultaneously
satisfy several predictions of quantum mechanics \cite{GHZ89}.
Although quantum nonlocality is intimately related to entanglement
\cite{VW02}, security of quantum cryptography \cite{Ekert91}, and
communication complexity \cite{BZPZ04}, there is as of yet no
loophole-free quantum nonlocality experiment. A particulary
important problem is the detection loophole. It occurs when the
imperfect efficiency of the detectors leaves room for LHV theories
in which undetected events can occur due to local hidden
instructions rather than to imperfections \cite{Pearle70}. An
appropriate measure of the quantum nonlocality of a given quantum
state and Bell inequality is, therefore, the minimum detection
efficiency required for a loophole-free Bell experiment, $\eta_{\rm
crit}$. It is defined as the value of the ratio between detected and
emitted particles such that, if $\eta \le \eta_{\rm crit}$, there is
a LHV theory reproducing the predictions of quantum mechanics, but
no such LHV theories exist if $\eta
> \eta_{\rm crit}$. The value of $\eta_{\rm crit}$ is known for some
scenarios \cite{GM87, Eberhard93, Larsson98a, Larsson98b, LS01,
CL07}, and some general bounds have been obtained \cite{Massar02}.
Curiously, $\eta_{\rm crit}$ was still unknown for a very important
scenario.

Eighteen years ago, Mermin and others discovered the first example
of a Bell inequality with a violation that grows exponentially with
the number $n$ of particles \cite{Mermin90a, Ardehali92}.
Specifically, they show that the $n$-qubit
Greenberger-Horne-Zeilinger (GHZ) state $|{\rm GHZ}_n\rangle$
\cite{GHZ89} violates a $n$-partite Bell inequality by an amount
that grows as $2^{(n-1)/2}$ \cite{Mermin90a, Ardehali92}. If instead
of a pure $|{\rm GHZ}_n\rangle$ we have a noisy one, $V |{\rm
GHZ}_n\rangle \langle {\rm GHZ}_n|+ (1-V) \openone /2^n$, then the
minimum value of $V$ required to observe a violation is $V_{\rm
crit}=2^{(1-n)/2}$. Later, Werner and Wolf proved that the Mermin
inequality is the two-setting correlation Bell inequality ``which
can be violated by the widest margin in quantum theory \ldots [and]
is the only one for which the maximal violation $2^{(n-1)/2}$ is
attained'' \cite{WW01}.

There have been several attempts to obtain $\eta_{\rm crit}$ for the
Mermin inequality: Braunstein and Mann showed that $\eta_{\rm crit}
\ge 2^{(1-n)/2n}$ for $n$ odd, and $\eta_{\rm crit} \ge
2^{(2-n)/2n}$ for $n$ even \cite{BM93}; Brassard, Broadbent, and
Tapp showed that $\eta_{\rm crit} < 2^{(2-n)/n}$ \cite{BBT05}; and
Larsson proved that $\eta_{\rm crit} = 3/4$ for $n=3$
\cite{Larsson98a}. However, no formula was known for arbitrary $n$
\cite{Larsson07}. In this Letter we prove that $\eta_{\rm crit} =
n/(2n-2)$. In addition, we obtain numerically the relation between
$\eta_{\rm crit}$ and $V_{\rm crit}$ for several values of $n$.

%%%%%%%%%%%%%%%%%%%%%%%%%%%%%%%%%%%%%%%%%%%%%%%%%%%%%%%%%%%%%%%%%%%

%\section{Definitions and notation}

%%%%%%%%%%%%%%%%%%%%%%%%%%%%%%%%%%%%%%%%%%%%%%%%%%%%%%%%%%%%%%%%%%%

The Mermin inequality is based on the GHZ proof of Bell's theorem
\cite{GHZ89}. It shows the impossibility of assigning predefined
values $-1$ or $1$ to local observables, simultaneously satisfying
several perfect correlations predicted by quantum mechanics. The
scenario for the GHZ proof is the following. A system composed of $n
\ge 3$ particles is initially prepared in the state $|{\rm
GHZ}_n\rangle$. Each particle moves away to a distant space-time
region where an observer measures randomly either $X_i$ or $Z_i$,
where $X$ and $Z$ denote the Pauli matrices $\sigma_x$ and
$\sigma_z$, and $i$ denotes particle $i$. Local measurements on
particle $i$ are assumed to be spacelike separated from the choices
of measurements made on all other particles.

The $n$-qubit GHZ state is the unique simultaneous eigenstate that
satisfies
\begin{equation}
g_i |{\rm GHZ}_n\rangle = |{\rm GHZ}_n\rangle,\;\;\text{ for
}i=1,\ldots,n, \label{erule}
\end{equation}
where
\begin{equation}
g_i = X_i \bigotimes_{j\neq i}^{n} Z_j \label{grule}
\end{equation}
are the generators of the stabilizer group of the GHZ state, defined
as the set $\{s_j\}_{j=1}^{2^n}$ of all products of the generators.
The perfect correlations of the GHZ state are
\begin{equation}
\langle {\rm GHZ}_n | s_j |{\rm GHZ}_n\rangle = 1,\;\;\text{ for
}j=1,\ldots,2^n. \label{perfectcorrelations}
\end{equation}

For $n$ odd, the GHZ proof is as follows. Any LHV theory assigning
predefined values $-1$ or $1$ to $X_i$ and $Z_i$ in agreement with
the quantum predictions given by (\ref{perfectcorrelations}) must
satisfy
\begin{equation}
s_j=1,\;\;\text{ for }j=1,\ldots,2^n. \label{qpredictions}
\end{equation}
However, if we take into consideration only the $2^{n-1}$
predictions (\ref{qpredictions}) involving stabilizing operators
$s_j$ that are products of an odd number of generators, and assume
predefined values either $-1$ or $1$, then it so happens that, at
most, only $2^{n-2}+2^{(n-3)/2}$ out of these $2^{n-1}$ predictions
are satisfied. For the remaining predictions of quantum mechanics,
the corresponding prediction of the LHV model is the opposite (i.e.,
$s_j=-1$); the reason for this behavior will be explained below [see
(ii)]. Therefore, this discrepancy between quantum mechanics and LHV
theories can be reformulated as a violation of a Bell inequality.
Any LHV theory must satisfy the following inequality:
\begin{equation}
|\beta_n| \le 2^{(n-1)/2}, \label{Merminodd}
\end{equation}
where the Bell operator
\begin{eqnarray}
\beta_n=\frac{1}{2}\left[\prod_{i=1}^n(\openone +
g_i)-\prod_{i=1}^n(\openone - g_i)\right] \label{Belloperator}
\end{eqnarray}
is the sum of all stabilizing operators which are products of an odd
number of generators. Inequality (\ref{Merminodd}) is the Mermin
inequality for $n$ odd \cite{Mermin90a}. On the other hand, the
$n$-qubit GHZ state satisfies
\begin{equation}
\langle {\rm GHZ}_n | \beta_n |{\rm GHZ}_n\rangle = 2^{n-1},
\label{Expectedvalue}
\end{equation}
and therefore violates the Mermin inequality (\ref{Merminodd}) by an
amount that grows as $2^{(n-1)/2}$ \cite{Mermin90a}. For $n$ even,
the Mermin inequality is not violated by $2^{(n-1)/2}$. The
equivalent inequality was found by Ardehali \cite{Ardehali92}. For
an explanation of the Ardehali inequality in terms of stabilizers of
the GHZ state, see \cite{GC08}. For simplicity's sake, we will focus
on the Mermin inequality for $n$ odd. Our proof works similarly for
$n$ even when we consider the Ardehali inequality.

Following \cite{Larsson98b}, we include the detector inefficiency in
the LHV model, so that the model consists of a set of instructions
telling the $n$ particles what to do if $X$ or $Z$ are measured. For
a given particle, the only possible instructions are ``give a
detection ($-1$ or $1$)'' or ``do not give a detection.''

$P(X_i)$ is the probability that particle $i$ is detected (giving
either $-1$ or $1$) when $X_i$ is measured. $P(X_i|X_j)$ is the
probability that particle $i$ is detected when $X_i$ is measured if
particle $j \neq i$ is detected when $X_j$ is measured. $P(X_i|X_j
Z_k)$ is the probability that particle $i$ is detected when $X_i$ is
measured if particle $j$ ($j \neq i$) is detected when $X_j$ is
measured and particle $k$ ($i \neq k \neq j \neq i$) is detected
when $Z_k$ is measured. Analogously, $P(X_i Z_j|Z_k)$ is the
probability that particle $i$ is detected when $X_i$ is measured and
particle $j$ ($j \neq i$) is detected when $Z_j$ is measured if
particle $k$ ($i \neq k \neq j \neq i$) is detected when $Z_k$ is
measured.

%%%%%%%%%%%%%%%%%%%%%%%%%%%%%%%%%%%%%%%%%%%%%%%%%%%%%%%%%%%%%%%%%%%

%\section{Properties of the LHV models}

%%%%%%%%%%%%%%%%%%%%%%%%%%%%%%%%%%%%%%%%%%%%%%%%%%%%%%%%%%%%%%%%%%%

In our LHV models, measurement results are predefined and are
independent of the measurements on other particles. In addition,
they must satisfy some restrictions dictated by the expected (and
testable) behavior of the detectors and the properties of the
$n$-qubit GHZ state for the measurements involved in a test of the
Mermin inequality. Specifically, the following assumptions lead to
the following restrictions.

(i) {\em All detectors have equal, constant detection efficiency.
The efficiency is the same when $X$ or $Z$ are measured. The
detection errors are independent. The detectors have no dark
counts.---}From these assumptions it follows that $P(A_i)=p$,
$\forall A \in \{X,Z\}$ and $\forall i \in \{1,2,\ldots, n\}$, and
$P(A_i,\ldots, B_j|C_k,\ldots, D_l)=p^r$, $\forall A,\ldots, B,
C,\ldots,D \in \{X,Z\}$ and $\forall$ (different) $i,\ldots,
j,k,\ldots, l \in \{1,2,\ldots n\}$; $r$ is the number of elements
in $A_i,\ldots,B_j$. That is, the different probabilities must be
symmetric under particle permutation and under the permutation of
$X_i$ and $Z_i$. If $p$ is the minimum over all possible LHV models,
then $\eta_{\rm crit}=p$.

(ii) {\em Compatibility with the statistical predictions of quantum
mechanics for the Mermin inequality using the $n$-qubit GHZ
state.---}Each of the terms obtained by expanding
(\ref{Belloperator}), e.g., $X_1 Z_2 Z_3\ldots Z_n$, represents an
experimental configuration required to test inequality
(\ref{Merminodd}). We will also consider configurations obtained for
the previous ones by selecting measurements on subsets containing an
odd number of particles, e.g., $X_1 Z_2 Z_3$. According to the
predictions of quantum mechanics for the $n$-qubit GHZ state, in
each of these experimental configurations, when an odd number $3 \le
q \le n$ of particles are detected, the corresponding results must
satisfy
\begin{eqnarray}
& X_i Z_j Z_k\cdots Z_q = Z_i X_j Z_k\cdots Z_q = Z_i Z_j X_k\cdots
Z_q = \cdots \nonumber \\ & = Z_i Z_j Z_k\cdots X_q =-X_i X_j
X_k\cdots X_q. \label{GHZnodd}
\end{eqnarray}
In addition, if $q \neq n$, then (\ref{GHZnodd}) must equal
$Z_{q+1}\ldots Z_n$. Therefore, depending on the result, $-1$ or
$1$, of the product $Z_{q+1}\ldots Z_n$, we can divide the reduced
state of the $q$ particles in two ensembles. For each of these
ensembles a different GHZ proof applies. If $q=n$, then
(\ref{GHZnodd}) must equal $1$. Since these conditions cannot be
fulfilled if $X$ and $Z$ of three or more particles have predefined
values either $-1$ or $1$, then we will conclude that the only
hidden instructions allowed in the LHV model are those in which $X$
and $Z$ of three different particles have not all of them predefined
values.

The challenge is to obtain the maximum possible detection efficiency
that can be reproduced with LHV models which satisfy (i) and (ii).
Each of these LHV models is defined on a probability space
$(\Lambda, \rho)$, and is made up of subsets of instructions
$I_{k,l,m}\subset \Lambda$, each of them characterized by three
numbers: $k$ is the number of particles for which both observables
($X$ and $Z$) are predefined (i.e., would give a detection when the
observable is measured), $l$ is the number of particles for which
only one of the observables ($X$ or $Z$) is predefined, and
$m=n-l-k$ is the number of particles for which none of the
observables are predefined.

%%%%%%%%%%%%%%%%%%%%%%%%%%%%%%%%%%%%%%%%%%%%%%%%%%%%%%%%%%%%%%%%%%%

%\section{Necessary condition}

%%%%%%%%%%%%%%%%%%%%%%%%%%%%%%%%%%%%%%%%%%%%%%%%%%%%%%%%%%%%%%%%%%%

Five lemmas are needed to prove our main result.

{\em Lemma 1.}---In order to find the maximum detection efficiency
that can be reproduced with LHV models which satisfy (i), it
suffices to consider LHV models where each of the subsets
$I_{k,l,m}$ satisfies (i).

{\em Proof.}---Suppose we find a LHV model compatible with a
detection efficiency $\eta$, such that some of the subsets
$I_{k,l,m}$ do not satisfy (i). Since the model must satisfy (i), we
can always symmetrize it in all possible ways (by changing $Z$'s to
$X$'s, and interchanging the different particles) and consider an
average of all these rearrangements. Clearly, this new model will
have the same $\eta$ and each of the $I_{k,l,m}$ will satisfy (i).
The new model will satisfy (ii) if and only if the original model
satisfied (ii).\hfill\endproof

Therefore, from now on we will only consider models such that each
of the $I_{k,l,m}$ satisfies (i). Each subset $I_{k,l,m}\subset
\Lambda$ occurs with probability $0 \le \rho_{k,l,m} \le 1$.

In order to satisfy (ii), the only subsets of instructions
$I_{k,l,m}$ allowed in our LHV models are those with $k=0,1,2$. In
addition, the predefined values must satisfy (\ref{GHZnodd}), and
the $-1$ and $1$ values must be suitably distributed in order to
reflect the fact that for the GHZ state all of the one qubit reduced
density matrices are maximally mixed. Notice that these last two
conditions are not particularly restrictive and can be easily
satisfied. Therefore, in order to improve the clarity of the
presentation, we will not insist on them hereafter.

An upper bound on $\eta$ will follow from probabilistic
considerations on each of the $I_{k,l,m}$. We will use the notation
$P_{I_{k,l,m}}$ to refer to the probabilities of detection of the
different variables within the sets $I_{k,l,m}$.

{\em Lemma 2.}---The value of $P_{I_{2,n-2,0}}(X_1|X_2,\ldots,X_n)$
(and all the possible substitutions of $X_i$ by $Z_i$ and
permutations of the indexes) is $n/(2n-2)$.

{\em Proof.}---By definition,
\begin{equation}
P_{I_{2,n-2,0}}(X_1|X_2,\ldots,X_n)=\frac{P_{I_{2,n-2,0}}(X_1,\ldots,
X_n)}{P_{I_{2,n-2,0}}(X_2,\ldots,X_n)}.
\end{equation}
In the subset $I_{2,n-2,0}$, only $\binom{n}{2}$ instructions have
predefined values for all the $X_i$'s. Since the total number of
different instructions in $I_{2,n-2,0}$ is $\binom{n}{2}2^{n-2}$,
then
\begin{equation}
P_{I_{2,n-2,0}}(X_1,\ldots,X_n)= \frac{1}{2^{n-2}}.
\end{equation}

In order to calculate $P_{I_{2,n-2,0}}(X_2,\ldots,X_n)$, we consider
the subset $S \subset I_{2,n-2,0}$ where both $X_1$ and $Z_1$ have
predefined values. We also consider the complementary subset
$S^c=I_{2,n-2,0}\setminus S$. Clearly,
$P_{I_{2,n-2,0}}(S)=\frac{\binom{n-1}{1}}{\binom{n}{2}}$ and
$P_{I_{2,n-2,0}}(S^c)=\frac{\binom{n-1}{2}}{\binom{n}{2}}$.
Reasoning in $S$ and $S^c$ as above, we see that
\begin{eqnarray}
& & P_{I_{2,n-2,0}}(X_2,\ldots,X_n) \nonumber \\
& & =P_{I_{2,n-2,0}}(X_2,\ldots,X_n|S)P_{I_{2,n-2,0}}(S) \nonumber
\\& &+P_{I_{2,n-2,0}}(X_2,\ldots,X_n|S^c)P_{I_{2,n-2,0}}(S^c)
\nonumber \\
& &=\frac{1}{2^{n-2}}\frac{\binom{n-1}{1}}{\binom{n}{2}}+
\frac{1}{2^{n-3}}\frac{\binom{n-1}{2}}{\binom{n}{2}}=\frac{2n-2}{n
2^{n-2}}.
\end{eqnarray}
\hfill\endproof

{\em Lemma 3.}---For every $I_{k,l,m}$ different than $I_{2,n-2,0}$,
$P_{I_{k,l,m}}(X_1|X_2,\ldots, X_n)$ is either undefined or less
than $n/(2n-2)$.

{\em Proof.}---If $m>1$, $P_{I_{k,l,m}}(X_1|X_2,\ldots, X_n)$ is not
defined. If $m=1$, $P_{I_{k,l,m}}(X_1|X_2,\ldots, X_n)=0$; hence, we
consider only the case $m=0$. In this case, there are only two
subsets to consider, $I_{1,n-1,0}$ and $I_{0,n,0}$.

Reasoning as in the proof of Lemma 2,
\begin{equation}
P_{I_{0,n,0}}(X_1|X_2,\ldots,X_n)=\frac{1}{2}
\end{equation}
and
\begin{equation}
P_{I_{1,n-1,0}}(X_1|X_2,\ldots,X_n)=\frac{n}{2n-1},
\end{equation}
which is always less than $\frac{n}{2n-2}$.\hfill\endproof

{\em Lemma 4.}---For efficiencies higher than $\eta = n/(2n-2)$,
there are no LHV models which simultaneously reproduce all the
quantum predictions (\ref{GHZnodd}) for $q$ odd and $3 \le q \le n$.

{\em Proof.}---The LHV model must satisfy
$\eta=P(X_1)=P(X_1|X_2,\ldots, X_n)$. The value of
$P(X_1|X_2,\ldots,X_n)$ must be less than or equal to the maximum of
the values, where defined, of $P_{I_{k,l,m}}(X_1|X_2,\ldots,X_n)$
for the different subsets $I_{k,l,m}$ of the LHV model. According to
Lemmas 2 and 3, all of these values are less than
$n/(2n-2)$.\hfill\endproof

%%%%%%%%%%%%%%%%%%%%%%%%%%%%%%%%%%%%%%%%%%%%%%%%%%%%%%%%%%%%%%%%%%%

%\section{Sufficient condition}

%%%%%%%%%%%%%%%%%%%%%%%%%%%%%%%%%%%%%%%%%%%%%%%%%%%%%%%%%%%%%%%%%%%

{\em Lemma 5.}---For $\eta = n/(2n-2)$, there are LHV models which
simultaneously reproduce all the quantum predictions (\ref{GHZnodd})
for $q$ odd and $3 \le q \le n$.

{\em Proof.}---We prove it by constructing explicit LHV models $M_n
(\eta)=\{(\rho_{k,l,m}, I_{k,l,m})\}$ reproducing the quantum
predictions for a given $n$ and $\eta$. Exact LHV models for
$n=3,4,5$ for $\eta=n/(2n-2)$ are the following:
\begin{widetext}
\begin{eqnarray}
& M_3 \left(\frac{3}{4}\right)= \left\{
\left(\frac{54}{64},I_{2,1,0}\right),
\left(\frac{9}{64},I_{1,0,2}\right),
\left(\frac{1}{64},I_{0,0,3}\right)\right\}, \\
& M_4
\left(\frac{2}{3}\right) = \left\{
\left(\frac{64}{81},I_{2,2,0}\right),
\left(\frac{8}{81},I_{2,0,2}\right),
\left(\frac{8}{81},I_{1,0,3}\right),
\left(\frac{1}{81},I_{0,0,4}\right)\right\}, \\
& M_5 \left(\frac{5}{8}\right) = \left\{
\left(\frac{25000}{2^{15}},I_{2,3,0}\right),
\left(\frac{3750}{2^{15}},I_{2,1,2}\right),
\left(\frac{1750}{2^{15}},I_{2,0,3}\right),
\left(\frac{2025}{2^{15}},I_{1,0,4}\right),
\left(\frac{243}{2^{15}},I_{0,0,5}\right)\right\},
\end{eqnarray}
\end{widetext}
where, e.g., $\left(\frac{54}{64},I_{2,1,0}\right)$
means that the model has instructions $I_{2,1,0}$ with probability
$\frac{54}{64}$, etc. For higher $n$, we have obtained LHV models
for $\eta=n/(2n-2)$ numerically for up to $n=15$ qubits.
For a given
$n$ and $\eta=n/(2n-2)$, the LHV models are not unique.

\begin{figure}[tbh]
\centerline{\includegraphics[bb= 50 100 750 550, clip, angle=0,
width= 1.02 \columnwidth]{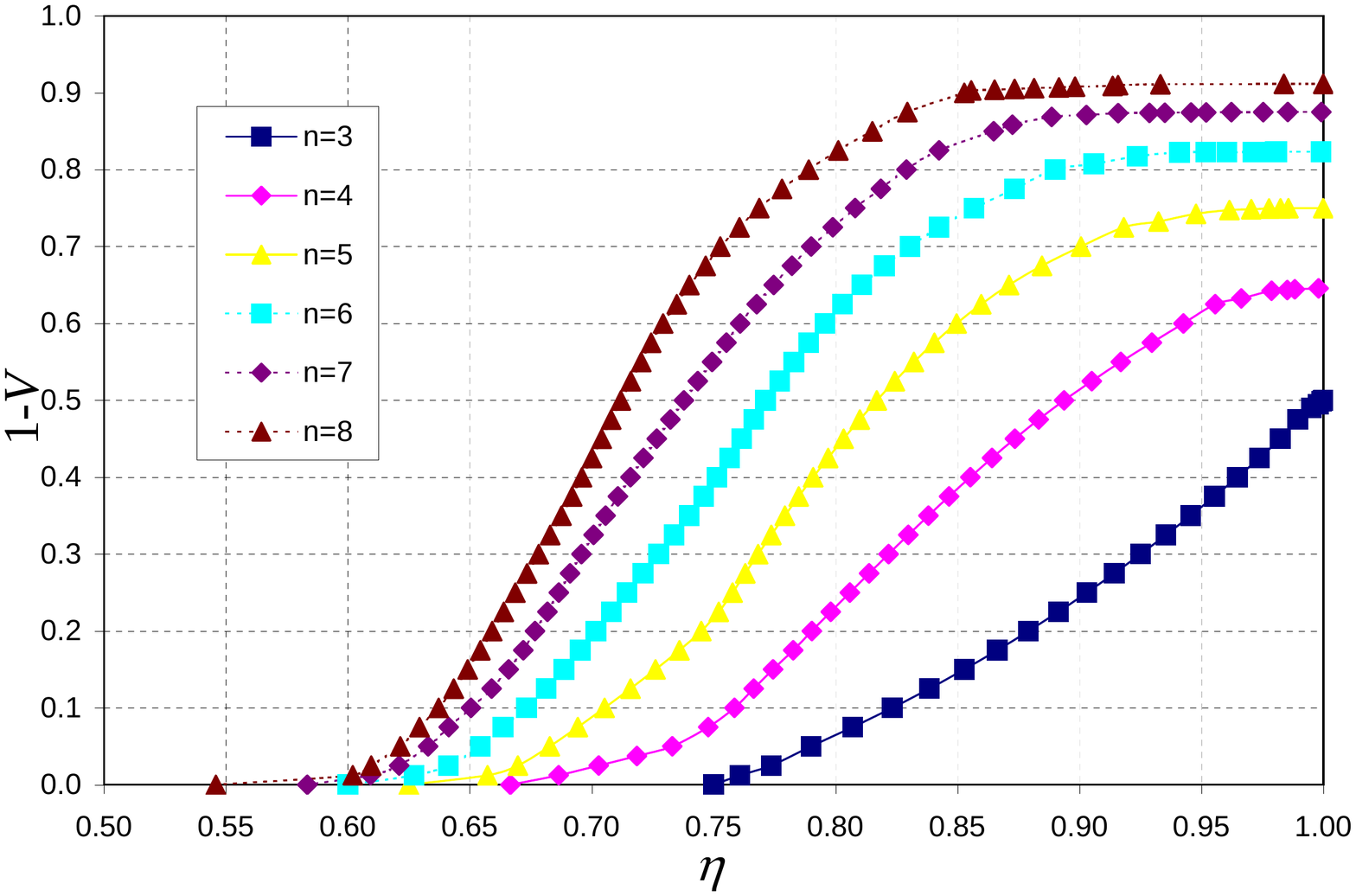}} \caption{\label{Fig1}
Maximum (background) noise $1-V$ as a function of the minimum
detection efficiency $\eta$ required to violate the Mermin
inequality, when the state is $V |{\rm GHZ}_n\rangle \langle {\rm
GHZ}_n|+ (1-V) \openone /2^n$, for $n=3,4,\ldots,8$ qubits. Note
that if $V=1$, then $\eta=n/(2n-2)$, and if $\eta =1$, then
$V=2^{(1-n)/2}$.}
\end{figure}

In addition, we have calculated numerically the maximum background
noise \cite{Eberhard93,CL07} as a function of the minimum detection
efficiency required to violate the Mermin inequality. The results,
for up to $n=8$ qubits, are summarized in Fig.~\ref{Fig1}.
%The detailed calculations will be presented elsewhere.

%%%%%%%%%%%%%%%%%%%%%%%%%%%% Figure 1 %%%%%%%%%%%%%%%%%%%%%%%%%%%%%

%%%%%%%%%%%%%%%%%%%%%%%%%%%%%%%%%%%%%%%%%%%%%%%%%%%%%%%%%%%%%%%%%%%

%\section{Conclusion}

%%%%%%%%%%%%%%%%%%%%%%%%%%%%%%%%%%%%%%%%%%%%%%%%%%%%%%%%%%%%%%%%%%%

We have proven that a loophole-free Bell experiment using an
$n$-qubit GHZ state and the correlations appearing in the
$n$-partite Mermin inequality requires a detection efficiency higher
than $n/(2n-2)$. This result solves a long-standing open problem and
is specially relevant for the $4$ \cite{SKKLMMRTIWM00}, $5$
\cite{ZCZYBP04}, and $6$-qubit GHZ states \cite{LKSBBCHIJLORW05}
prepared in recent experiments. $n/(2n-2)$ is the threshold
efficiency beyond which there is no LHV model which simultaneously
satisfies all the quantum predictions (\ref{GHZnodd}) and is the
critical efficiency beyond which there is no LHV model reproducing
all the quantum predictions for all the Bell inequalities, with two
settings for $q$ observers ($3 \le q \le n$) and one setting for the
other $n-q$ observers, contained in the Mermin inequality. This
observation is of practical interest since, e.g., testing each of
the $\binom{3}{5}$ $2$-$2$-$2$-$1$-$1$-setting Bell inequalities on
a $5$-qubit GHZ state requires only $3$ spacelike separated regions,
while testing the $5$-partite Mermin inequality requires $5$
spacelike separated regions. When $n$ tends to infinity, $\eta_{\rm
crit}$ tends to $1/2$, reflecting the fact that $I_{0,n,0}$ is a
trivial LHV model compatible with the quantum predictions if $\eta
\le 1/2$.

%%%%%%%%%%%%%%%%%%%%%%%%%%%%%%%%%%%%%%%%%%%%%%%%%%%%%%%%%%%%%%%%%%%

%\section*{Acknowledgments}

%%%%%%%%%%%%%%%%%%%%%%%%%%%%%%%%%%%%%%%%%%%%%%%%%%%%%%%%%%%%%%%%%%%

\begin{acknowledgments}
The authors thank J.-\AA. Larsson and O. G{\"u}hne for useful
conversations. A.C. acknowledges support from the MEC Project No.
FIS2005-07689, and the Junta de Andaluc{\'\i}a Excellence Project
No. P06-FQM-02243. I.V. acknowledges support from the MEC Project
No. MTM2005-00082.
\end{acknowledgments}

%%%%%%%%%%%%%%%%%%%%%%%%%%% References %%%%%%%%%%%%%%%%%%%%%%%%%%%%

%%%%%%%%%%%%%%%%%%%%%%%%%%%%%%%%%%%%%%%%%%%%%%%%%%%%%%%%%%%%%%%%%%%

\end{document}